 \documentclass{emulateapj}
 \usepackage{times}
 \usepackage{lscape}

\def\la{\mathrel{\mathpalette\fun <}}
\def\ga{\mathrel{\mathpalette\fun >}}

\def\fun#1#2{\lower0.837ex\vbox{\baselineskip0ex\lineskip0.209ex
  \ialign{$\mathsurround=0ex#1\hfil##\hfil$\crcr#2\crcr\sim\crcr}}}

\def\msun{M_\odot}
\def\msunyr{M_\odot \ {\rm yr}^{-1}}

\def\msuns{M_\odot \ {\rm s}^{-1}}
\def\sles{\lower2pt\hbox{$\buildrel {\scriptstyle <}
   \over {\scriptstyle\sim}$}}

\def\sgreat{\lower2pt\hbox{$\buildrel {\scriptstyle >}
   \over {\scriptstyle\sim}$}}

\def\la{\mathrel{\mathpalette\fun <}}
\def\ga{\mathrel{\mathpalette\fun >}}

\begin{document}

\title{A New Paradigm for Gamma Ray Bursts:
       Long Term Accretion Rate Modulation by an External Accretion Disk}
\shortauthors{CANNIZZO \& GEHRELS}
\author{
J.~K.~Cannizzo\altaffilmark{1,2},
N.~Gehrels\altaffilmark{2}
}
\altaffiltext{1}{CRESST/Joint Center for Astrophysics,
                 University of Maryland, Baltimore County, Baltimore, MD 21250,
                 John.K.Cannizzo@nasa.gov}
\altaffiltext{2}{NASA-Goddard Space Flight Center, Greenbelt, MD 20771,     
                 gehrels@milkyway.gsfc.nasa.gov}

\begin{abstract}
We present\footnote{submitted to the 
Astrophysical Journal: 22 Jan 2009; accepted: 23 May 2009}
  a new way of looking at the very long term evolution
of GRBs in which the disk of material 
surrounding the putative black hole powering the GRB jet
modulates the mass flow, and hence the efficacy of the
process that extracts rotational energy from the black hole
  and inner accretion disk.
The pre-{\it Swift} paradigm of 
  achromatic,  shallow-to-steep
  ``breaks''
   in the long term GRB light curves has not been 
borne out by detailed {\it Swift} data 
amassed in the past several years. 
We argue that, given the initial existence of a fall-back disk
near the progenitor, 
  an unavoidable consequence will be 
  the formation of an ``external disk''
  whose outer edge
   continually moves to larger radii
    due to 
   angular momentum transport
    and lack of a confining torque.
The mass reservoir at
large radii moves outward with time
and gives a natural power
law decay to the GRB light curves. In this model, the different
canonical power law decay segments
in the GRB identified by Zhang et al.
    and Nousek et al.
    represent different
physical states of the accretion disk. We identify a physical disk
state with each power law segment. 
\end{abstract}


\section{Introduction}

Gamma-Ray Bursts (GRBs) are thought to be energetic events
   heralding 
  the formation of black holes (BHs):
 the explosion of 
   a massive star  at cosmological distances
  gives rise to a long GRB, or two neutron
stars coalesce to make a short GRB.
  (For a recent review of GRBs,
   see Gehrels, Ramirez-Ruiz, \& Fox 2009.)
  To see a GRB at all,
  our line-of-sight must lie close to the BH angular momentum 
  vector.  
A jet  
  is launched toward 
   the observer as the BH forms.  
  The initial Lorentz factor of material 
   in the jet $\sim 10^2 - 10^3$
is large enough so that the observer only sees the central portion
of the jet, due to relativistic beaming. 
The transition from ``spherical'' to ``jet-like'' expansion of
the GRB ejecta is the basis of the current framework
within which the analysis of GRB light curves
 is carried out (Rhoads 1997, 1999; Sari, Piran, \& Halpern 1999).
  Parts of
the jet further away from the symmetry axis are 
  initially  invisible to the
observer, hence the expansion is indistinguishable from that of
of a relativistically expanding sphere.
 Eventually
interaction with the circumstellar medium
reduces the Lorentz factor until the edges of
the jet are visible, after which the decay of the GRB afterglow
light curves becomes more rapid.
        This fundamental property of the decay relies only on
special relativity and is not specific to any one
wavelength so these theoretical decays should have a similar, strong
``break'' from shallow to steep across  all wavebands, hence the
designation ``achromatic break''.

Frail et al. (2001) studied the decay properties of 
17 pre-{\it Swift} GRBs with redshifts. By analyzing the multiwavelength
decay properties, they were able to determine putative break times,
which, when combined with distances and 
isotropic-equivalent GRB energies,
allowed a correction to the true, beaming-corrected energy.
Frail et al. (2001, see their Fig. 2) show how a spread in
isotropic energies between $\sim 10^{52}$ and  $\sim 10^{54}$ erg 
       for their sample  
       is considerably reduced down to a standard,
geometry-corrected energy centered at $\sim3\times 10^{50}$ erg.
 A  detailed examination of the data entering into their study
(see their Table 1) shows the sparse
  coverage of many of the 
 light curves 
 and underscores the difficulty
  in reliably assigning
 break times\footnote{The 
data used by Frail et al., as well as that for all GRBs, is viewable at
{\tt http://grblog.org/grblog.php}}. 
An updated study along the same
  lines as  Frail et al. (2001) by 
   Bloom, Frail, \& Kulkarni (2003)
  contains a somewhat larger sample $-$
28 GRBs with redshifts.
  We should emphasize that we 
   do not argue against the
need for beaming; clearly energies as large as $\sim 10^{54}$ erg
 are problematic, given that $\msun c^2 = 1.8 \times 10^{54}$ erg.
  It is important to distinguish,
   however,  between
 a time-variable relativistic
  beaming, and geometric beaming.
 In this work
we adopt a time-constant geometric beaming
   factor $f_{\rm beam} = 3\times 10^{-3}$
  to convert from isotropic energy to true energy.
  The question of interest to us is whether the very long
term decay in the flux from GRBs, particular the X-ray flux,
is due to the deceleration of a jet, or a decreasing mass
supply rate onto the central engine.

It was anticipated
that {\it Swift} (Gehrels et al. 2004)
   would provide good examples of shallow-to-steep achromatic decays. 
After
$\ga400$  {\it Swift} GRBs, and now a total of $>170$ redshifts
for all GRBs (41 for pre-{\it Swift} GRBs), this
has not turned out to be the case. 
 In spite of many GRBs with exquisite
wavelength coverage, observed over many $e-$foldings of decay in flux,
  there is only one GRB
  in the {\it Swift} era 
   perhaps
   showing an achromatic
  break, seen in X-ray and optical
   (GRB060206 - Curran et al. 2007)
and a small number  with
candidate jet-breaks
(Blustin et al. 2006,
  Stanek et al. 2007,
  Dai et al. 2007,
   Willingale et al. 2007,
   Kocevski \& Butler 2008);
  the overwhelming majority
   do not show one 
   (e.g., 
    Oates et al. 2007;
         Racusin et al. 2008;  Liang et al. 2008).
  Liang et al. (2008) present a thorough
  investigation of {\it Swift} XRT and optical
  jet break candidates.
  They note that  ``It is
 fair to conclude that we {\it still}
  have not found a textbook
  version of a jet break after many years of intense
observational campaigns.''

   Zhang et al. (2006, see their Fig. 1;
  also Nousek et al. 2006, see their Fig. 3)
   present a simple schematic for
the decaying GRB light curve as seen by the XRT on {\it Swift}.
The decay is traditionally shown in $\log F - \log t$.
There are four basic power-law decay ($F \propto t^{-\alpha}$)
regimes: (I) $\alpha_{\rm I} \simeq  3$   out to $10^2-10^3$ s,
        (II) $\alpha_{\rm II} \simeq  0.5$ out to $10^3-10^4$ s,
       (III) $\alpha_{\rm III} \simeq  1.2$ out to $10^4-10^5$ s,  and
        (IV) $\alpha_{\rm IV} \simeq    2$ at late times.
The general picture given in Zhang et al.
based on a handful of case studies from early {\it Swift} XRT 
results has been borne out by studies relying on a much larger sample
(e.g., O'Brien et al. 2006, 
  Ghisellini et al. 2009).  
In addition one also frequently sees flares within segment II 
(Burrows, et al. 2005;
   for possible explanations see
  Perna, Armitage, \& Zhang 2006,
  Lazzati, Perna, \& Begelman 2008),
which are beyond the scope of this work.
Figure 1 shows a schematic reproduction
of the idealized Zhang et al. light curve, 
  with realistic numbers for
the XRT flux.
  The conversion from the XRT flux values to
 a beaming-corrected luminosity
   is given by
   $L_{\rm XRT} = 4\pi d_L^2 f_{\rm beam} F_{\rm XRT}$,
  where $d_L$
  is the luminosity distance.

\begin{figure}
\centering
\epsscale{1.0}
\plotone{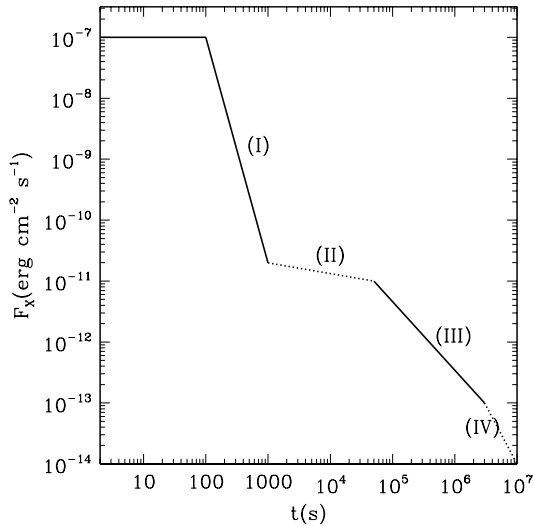}
\vskip -2cm
\figcaption{
A schematic 
representation of the X-ray decay of a
{\it Swift} GRB
as seen with XRT, based on the 
Zhang et al. (2006) idealization
and GRB 060729 from Willingale et al. (2007,
  see their Fig. 10).
   To account for cosmological time
dilation for a given GRB,
    the $x-$axis needs to be shortened
by a factor $1+z$.
\label{fig1}}
\smallskip
\end{figure}

In this work we present a framework based on the 
long term evolution of a transient accretion disk
  formed out of some small fraction 
  of the progenitor
to account for segments II, III, and IV.
 (Segment I is too steep to be accounted for in 
our formalism.)
   The flatness of segment II arises from the
transient adjustment of a small amount
of gas into an accretion disk, the segment III
  decay is consistent with the self-similar decay of
a disk without a confining external torque, and the
increase in slope in segment IV would be caused by 
the onset of significant disk evaporation (or by the late-time
      accretion from a super-Eddington slim disk).
  The consistency of the theoretical decay with segment III
   argues for a direct association, and the fact that
it takes some time for an arbitrary initial mass profile
to adjust into a disk leads to the transient interval (segment II).

This work applies mainly to 
the commonly observed ``long'' GRBs
which  are 
believed to be due to
 the 
core collapse 
of 
a massive, high angular momentum progenitor, 
leading to the formation of black hole  $-$ 
 the ``collapsar'' model (Woosley 1993; 
MacFadyen \& Woosley 1999; Woosley \& Bloom 2006;
   Nagataki et al. 2007).
  There follows a period of electromagnetic extraction of
energy from the BH which  powers
 a 
 collimated jet.
For $\sim 10^2 - 10^3$s
following the phase of prompt GRB emission,
  $\sim0.1\msun$
   of bound matter
   from the disrupted  progenitor
 star returns to the BH
vicinity  
   due to conservation of angular momentum.
 Kumar, Narayan, \& Johnson (2008ab)
  attempt  to explain  the long-term 
{\it Swift}
X-ray light curves in terms of various 
phases of an accretion model. 
  They argue that the rapid accretion of the
outer half of the core of the progenitor,
possessing a density profile that drops 
steeply with radius, can lead to the initial, rapid decay.
   For the flat segment 
   (II) they
  discuss briefly the idea of having a small
  accretion  disk viscosity, but ultimately
discard  this explanation in favor of prolonged
fall-back of gas from the envelope of the progenitor at larger radii,
   which they argue can supply 
  matter to the central engine at
a near-constant rate, or a rate that
  decreases slightly with time.
   Kumar et al. (2008a)
    also present analytic time dependent 
  solutions for an accretion disk in 
  an ADAF-like state
(ADAF = advection dominated accretion flow)
and show that, for a radial mass profile ${\dot M}(r)\propto r^s$,
   the decay law would be
   $\alpha_{\rm III} =  4(s+1)/3$.
   For $s=0$, i.e., no wind mass-loss from the disk,
   the decay 
         is close to the Zhang et al. (2006) value.
    Under the assumptions of core fall-back
   for segment I and envelope fall-back for segment II,
   Kumar et al. (2008b) place constraints on the
  progenitor core and envelope radii, their density profiles, 
  and the pre-SN rotation rates of the core and envelope,
  normalized to their Keplerian values.

In Section 2 we discuss the physics of accretion disks, both steady
state and time dependent. 
  Section 3 presents a scenario for the overall, 
long term evolution of the external accretion disk,
  section 4 presents a discussion,
  section 5 discusses the implications of the theory, 
 and section 6 sums up.

\section{Accretion Disk Physics}

Most of the
previous discussion of GRB fall-back disks has failed to
recognize one unavoidable aspect of their existence:
namely, their   outer edge moves outward with time, carrying
with it mass and angular momentum.
   (Notable exceptions are Kumar et al. [2008ab].)
One can make simple arguments about  the 
perturbation accretion time scale
near the central engine based on physical conditions
there, but the outer portions are causally linked to the inner
regions through angular momentum transport,
and it is these
   outlying regions that 
  set the long-term accretion disk time scales.

   The physical agent responsible
   for angular momentum transport
          and
turbulent dissipation of energy
  is now believed to be the magneto-rotational
instability (MRI) or Balbus-Hawley instability (Balbus \& Hawley 1998)
which involves the shearing amplification of a weak seed 
magnetic field.
   The general process of angular momentum transport
  and energy dissipation in accretion disks is
    still commonly referred to as ``viscosity''.
 Balbus \&  Papaloizou (1999)
    show a foundation  for the dynamical ``$\alpha$''
  view of accretion disks
  (Shakura \& Sunyaev 1973)
   within the MRI.

 Extensive study of accretion disks in interacting binaries
   has guided thought on accretion disks
in other types of systems, such as around BHs in AGNs. 
 For instance,  the best constraint
we have of the magnitude of the $\alpha$ parameter comes from
a detailed study of the rate of decay of dwarf nova outbursts (Smak 1984),
and the values obtained for ionized gas
   $\alpha \simeq 0.1-0.2$ have been
carried over to accretion disks in other systems.

One aspect of accretion disks which is quite different 
for
isolated BHs is the lack of a constraint on the outer radius.
The sudden introduction of
   an annulus of gas
       near a central mass leads to
a process of smearing of the matter 
    into a broadened  ring. The ring  spreads
both to smaller and larger radii in response 
       to outward transport of angular 
momentum. As mass drains from the inner edge, 
an ever smaller amount
of material moves to larger and larger radii.
Pringle (1974)   studied ``external accretion disks''
in which matter initially in a torus spreads both to smaller and larger radii.
The class of solutions he considered of relevance to 
  this work
is that for disks with constant total angular momentum and decreasing mass.
Pringle (1991) generalized his earlier results.

By writing the equations for mass continuity 
and angular momentum transport
   in cylindrical coordinates,
  assuming 
 Keplerian rotation  $\Omega_K^2=GM_{\rm BH} r^{-3}$
   and
 a thin accretion disk,
   and
integrating  over the vertical thickness of the accretion disk, 
one arrives at an equation for the evolution of the surface density
$\Sigma=2\rho h$, where $\rho$ is the density and $h$ 
the disk semithickness (actually pressure scale height),
\begin{equation}
{\partial  \Sigma  \over \partial t} =
{3\over r} {\partial  \over \partial r}
\left[ r^{1/2} {\partial  \over \partial r} \left( \nu \Sigma r^{1/2} \right)\right].
\label{evolution}
\end{equation}
 The kinematic viscosity coefficient
\begin{equation}
\nu = {2 \alpha P  \over 3 \Omega_K \rho},
\end{equation}
where $P$ is the pressure and $\alpha$ is the Shakura-Sunyaev parametrization
of the angular momentum transport and heating (Shakura \& Sunyaev 1973).
  Debate persisted over many years as whether the viscosity should
scale with the total pressure or gas pressure. A scaling with the total pressure
leads to a strong dynamical instability 
in regions where $P_{\rm rad} > P_{\rm gas}$ $-$
the Lightman-Eardley instability (Lightman \& Eardley 1974)
    $-$
  which manifests itself as a limit cycle operating 
 in
 the inner, radiation pressure dominated portions of the disk
  (e.g., Cannizzo 1997).
However, the fact that           X-ray binaries
with very high rates of accretion, such as Sco X-1,
   often have perfectly stable light curves
leads one to conclude
that the Lightman-Eardley instability is 
  probably the artifact of an incorrect assumption
  for thin disks,
  namely $\nu \propto P_{\rm total}$ 
(Done, Gierli\'nski,  \& Kubota 2007).
  With the advent of the MRI, detailed MHD calculations
show that the saturation limit of the instability 
 depends
on the magnetic energy density 
  reaching a limit that may
depend on surface density
rather than pressure 
 (e.g., Hirose, Krolik, \& Blaes 2009).
  Hirose et al.
find not only that radiation dominated disks are thermally stable,
but that 
     fluctuations in magnetic field strength
   causally precede those in
  pressure (both radiation and gas). 
      Hirose et al. suggest adopting $\Sigma$ 
as the relevant independent variable.

The MRI couples linearly to the Keplerian shear
in the disk and dissipates energy through turbulent transport.
In formulating the steady state
  equations for the radial profile
 of accretion disk structure,
one begins by equating a heating function
\begin{equation}
Q_{\rm vis}^+ =  {9\over 8} \nu \Sigma \Omega_K^2
\end{equation}
  with a cooling
function 
\begin{equation}
Q_{\rm rad}^- = { 4\over 3} {ac \over \kappa \rho} {T^4 \over h},
\label{rte}
\end{equation}
assuming 
 an optically thick vertical structure
with  energy  transported 
   by
 radiative diffusion.
  The vertically integrated stress
provides the total radiated flux
  (assuming no radial advection
  of energy)
\begin{equation}
Q = \sigma T_{\rm eff}^4 = {3 \over 8\pi}  {\dot M} \Omega_K^2.
\label{Q_total}
\end{equation}
Finally, the condition of vertical hydrostatic
equilibrium  in the disk gives
\begin{equation}
{P_{\rm total}\over \rho} = \Omega_K^2 h^2.
\end{equation}

\subsection{Blandford-Znajek Coupling to the Inner Disk }

After a period of equilibration and 
   expansion, the external disk  may span
many decades in radius $-$
 from  $\sim 10^7$ to $\sim 10^{11}$ cm.
  The physical state of the disk
   is expected to vary enormously as one progresses
from small to large radii.
 In terms of  mediating the flow of mass down onto
the central engine, the similarity solutions just discussed
rely on the physical state of the outer disk.
  The physical state of the inner disk  determines the 
 strength of the Blandford-Znajek (BZ) process
  (Blandford \& Znajek 1977)
that extracts rotational energy from the 
central BH.
Ghosh \& Abramowicz (1997)
  examined the dependence of the
strength of the BZ mechanism on
physical conditions in the inner disk. 
The energy density of the
  magnetic field entrained within the disk
 sets the BZ luminosity, and insofar as the saturation
limit is set by equipartition with the pressure, 
the maximum pressure $p_{\rm max}$
   is of relevance.

Livio, Ogilvie, \& Pringle (1999)
   investigated the role of
 the BZ mechanism in  powering outflows
  and found that if the magnetic field
    threading the inner disk does not differ
  significantly in strength from that threading the 
  black hole,
    the electromagnetic output from the inner
  disk regions should dominate over that from the hole.  
   In simple terms, 
 the disk is more efficient
  at holding and accelerating the magnetic field
because its magnetic diffusivity 
is much less than that of the 
  BH.

However, more recent works using sophisticated GRMHD
  codes to follow the detailed magnetohydrodynamic launching of flows 
  from the very inner disk 
   find them to be highly efficient
  in terms of tapping the available supply 
   of rest mass-energy in the disk.
(McKinney \& Narayan 2007ab; Krolik, Hawley,  \& Hirose 2007).
  McKinney \& Narayan (2007a)
  note that previous simple estimates of the BH and disk
   failed to take into account the conversion of EM
power into thermal and material power in the 
corona and disk wind (Ghosh \& Abramowicz 1997, 
  Livio, Ogilvie, \& Pringle 1999)
  and so may have seriously overestimated
   the power from the disk,
  in which case the BH power may dominate.
   Krolik, Hawley,  \& Hirose (2007, see their Table 1)
   find high efficiencies $\epsilon_{\rm beam}$
   for the jet power, measured either in terms of 
 mass, radiative energy, or EM energy,
    normalized to the
         rest mass-energy accretion in the disk ${\dot M}c^2$.
  Their radiative
  efficiency values 
  (which they denote $\eta_{\rm NT}$)
  increase with $a/M$,
  ranging from $0.057$ for  $a/M=0$ up to $0.264$
   for $a/M=0.99$.
   The efficiencies may increase with time due to 
   BH spin-up.
   These workers find the jet  to be Poynting flux 
   dominated near the BH spin axis, giving strength
   to the idea of a lightweight jet that responds quickly
   to changes in ${\dot M}$ in the inner disk.
   In this work we adopt $\epsilon_{\rm beam}=0.1$
   as a fiducial value
   relating the jet energy production 
   as seen in X-rays to the accretion
   rest mass-energy within the main body of the accretion disk.

Initially the accretion rate can 
approach $\sim 1\msuns$, 
i.e., 
  super-Eddington 
  by a factor $\sim10^{14}$.
  One might  question 
   whether an accretion
    disk description 
  is fully applicable.  
  Studies
  of the inner disk
  at these very early 
times find
  that
   the temperatures and densities are so large that
the cooling is by neutrino emission
 (Narayan, Piran, \& Kumar 2001,
  Kohri \& Mineshige 2002, 
  Kohri, Narayan, \& Piran 2005,
Chen \& Beloborodov 2007).
Chen \& Beloborodov (2007)
show   that for a BH with high spin $a\sim1$
a neutrino cooling dominated  disk  
 transitions to an advective disk
at $\sim 10^{-3} - 10^{-2}\msuns$.
Kohri \& Mineshige (2002, see their Fig. 1)
detail the different physical regimes of these disks.
The cooling function
\begin{equation}
Q^{-} = Q_{\rm rad}^{-} + Q_{\rm adv}^{-} +  Q_{\rm \nu}^{-},
\end{equation}
where 
  $Q_{\rm \nu}^{-}$ is the neutrino cooling.

\subsection{ Eddington Luminosity and Accretion Rate}

For the GRB depicted schematically in Fig. 1, 
    $z=0.54$ and $d_L=3.1$ Gpc.
The inferred, 
  beaming-corrected X-ray luminosity
 corresponding to the earliest time
we seek to address, namely the onset of segment II, 
 would be  $L_{\rm II}
=4\pi d_L^2 f_{\rm beam} F_{\rm XRT, \ II}
  \simeq 3\times 10^{43}$ erg s$^{-1}$,
 where we used $F_{\rm XRT, \ II}=10^{-11}$ 
   erg cm$^{-2}$ s$^{-1}$ 
   (Grupe et al. 2007; see their Fig. 4)
    and  $f_{\rm beam} = 3\times 10^{-3}$.
  The isotropic-equivalent
   X-ray plateau luminosity for 
this GRB, $\sim 10^{46}$ erg s$^{-1}$,
   is in about the middle 
   (logarithmically)
  of the range $10^{44}$ erg s$^{-1}$ 
    $\la 
   4\pi d_L^2  F_{\rm XRT, \ II} 
     \la  10^{48}$ erg s$^{-1}$
  shown in 
  Dainotti et al. (2008; see their Figs. 1 and 2).
 Adopting  
    $\epsilon_{\rm beam} = 0.1$,
    the geometry-corrected 
    luminosity translates into a rate of accretion
in the inner disk  $L_{\rm II}/(\epsilon_{\rm beam}c^2) \simeq 5.3 \times 10^{-3} \msunyr$.
 In  the conventional view,
 for rates of accretion larger
  than the Eddington rate
  \begin{equation}
 L_E = {4\pi G M_{\rm BH} c \over \kappa_{\rm es} }
   \simeq 1.25 \times
10^{39}(M_{\rm BH}/10\msun) \ {\rm erg \ s}^{-1}
\end{equation}
  radiation pressure impedes the accretion process
  (see below, however, for the possibility of a slim
disk).
   The Eddington accretion rate  depends on 
  radius, through the local efficiency with which gravitational
energy is radiated $\epsilon_{\rm acc}(r) \equiv GM_{\rm BH}/(r c^2)$.
The locally defined Eddington accretion rate
  ${\dot M}_E(r)$
    is related to the Eddington luminosity
 via
 \begin{equation}
 L_E = \epsilon_{\rm acc}(r) {\dot M}_E(r) c^2.
\end{equation}
  Thus $\epsilon_{\rm acc}(r) \propto r^{-1}$ and  ${\dot M}_E(r)\propto r$,
   varying from $1.3\times 10^{-7}\msunyr$
 in the inner disk $6GM_{\rm BH}/c^2$
to $1.5 \times 10^{-3}\msunyr$ at $10^{11}$ cm.
Therefore in the region of interest to us in this study, 
  namely the outermost part of the disk, 
  the rate would only be 
  locally super-Eddington by a factor $\sim3$
  at this early time.    
 
  The preceding discussion was based 
  on the standard
assumption of  solar composition material, which is 
almost certainly not the case here.
   The fall-back matter comprising the disk
   would be expected 
  to be dominated by 
   $r-$processed
   material - Ni, Si, and Fe group elements.
   Calculations of the Eddington limit for such material
indicate   a potential suppression of the standard Eddington
 value by up to about three orders of magnitude 
  due to increased opacity from lines
 (e.g.,  Fryer, Colgate, \& Pinto 1999).
  However, this applies mainly to much lower temperatures
  than those discussed in the next section.

\subsection{Steady State Physics}

\subsubsection{ Standard Thin Disk}

 Cannizzo \& Reiff (1992)
  present general Shakura-Sunyaev scalings for
disks with arbitrary opacity law.
  For optically thick disks for which
  $\nu\propto P_{\rm gas}$,
$\kappa = \kappa_{\rm es} = 0.2(1+X) = 0.2$ 
  cm$^2$ g$^{-1}$ (no hydrogen),
and mean molecular weight $\mu=16$   
 (appropriate for the expected for mixture 
 of alpha elements - 
   predominantly 
  $^{56}$N, $^{28}$Si,$^{16}$O, and $^{12}$C
  - Fryer, Colgate, \& Pinto 1999),
the surface density
  \begin{equation}
\Sigma = 2.84 \times 10^7 \ {\rm g} \ {\rm cm}^{-2} 
         \  r_{11}^{-3/5}
         \  m_{{\rm BH}, \ 1}^{1/5}
         \  \alpha_{-1}^{-4/5}
         \ {\dot m}_{-2}^{3/5},
\label{std_sigma}
\end{equation}
  where  $r_{11} =r/10^{11}$ cm,
        $  m_{{\rm BH}, \ 1} = M_{\rm BH}/10\msun$,
         $ \alpha_{-1} = \alpha/0.1$,
          ${\dot m}_{-2} = {\dot M}/10^{-2}\msunyr$.
  (In the region of interest 
  for this study ${\dot m}_{-2} \simeq 0.01-1$.)
The midplane temperature
 \begin{equation}
 T     = 7.83 \times 10^6 \ {\rm K}
         \  r_{11}^{-9/10}
         \  m_{{\rm BH}, \ 1}^{3/10}
          \  \alpha_{-1}^{-1/5}
         \ {\dot m}_{-2}^{2/5}.
\end{equation}
  The disk temperature is  
  less than 
       the virial temperature
 \begin{equation}
T_{\rm virial} = {1\over 6} {\mu \over {\mathcal R}} {GM \over r} 
         = 4.26 \times 10^8 \ {\rm K}
         \  r_{11}^{-1}
         \  m_{{\rm BH}, \ 1}.
\end{equation}
  The only assumptions needed up to this point
are $Q^+ = Q^- = Q$.
  To obtain the disk scale height and density, we 
now specify $P=P_{\rm gas}$.
   This yields
a pressure scale height 
to disk radius ratio
 \begin{equation}
 {h \over r}  = 0.055
         \  r_{11}^{1/20}
         \  m_{{\rm BH}, \ 1}^{-7/20}
          \  \alpha_{-1}^{-1/10}
         \ {\dot m}_{-2}^{1/5},
\end{equation}
      a density
\begin{equation}
\rho =   2.57 \times 10^{-3} \ {\rm g} \ {\rm cm}^{-3} 
         \  r_{11}^{-33/20}
         \  m_{{\rm BH}, \ 1}^{11/20}
          \  \alpha_{-1}^{-7/10}
         \ {\dot m}_{-2}^{2/5},
\end{equation}
  and self-gravity to central object 
gravity
 \begin{equation}
 {g_s \over g_c }
   = {2\pi G\Sigma \over {\Omega_K^2 h}  }  
   =  1.62\times 10^{-3} 
         \  r_{11}^{27/20}
         \  m_{{\rm BH}, \ 1}^{-9/20}
          \  \alpha_{-1}^{-7/10}
         \ {\dot m}_{-2}^{2/5}.
\end{equation}

     The radiation pressure
     to gas pressure ratio
\begin{equation}
  {P_{\rm rad}  \over P_{\rm gas}  } 
           = 91
         \  r_{11}^{-21/20}
         \  m_{{\rm BH}, \ 1}^{7/20}
          \  \alpha_{-1}^{1/10}
         \ {\dot m}_{-2}^{4/5}
\end{equation}
   exceeds unity, therefore the pressure scale height,
  which was predicated upon $P=P_{\rm gas}$,
  should be increased by a factor $\sqrt{92} \approx  10$
  to $h/r\simeq 0.5$
  because of hydrostatic equilibrium.
   Note that 
   because
   we  adopt  $\nu\propto P_{\rm gas}$,
   the fact that
   $P_{\rm rad} > P_{\rm gas}$
   is not problematic
  in itself 
  as long  as $h/r<<1$.  
 However, the fact that  $h/r\simeq 0.5$
   begins to be problematic.

 As a last consistency check,
       the ratio of the flux of radially advected 
  energy $({\dot M}/2\pi r^2)c_s^2$  
   to vertically transported energy
\begin{equation}
 {Q_{\rm adv}^- \over Q_{\rm rad}^- }
   = { 3 \over 16 \pi}
       { \kappa_{\rm es}  {\mathcal R}
         \over
      ac \mu } 
      { {\dot M} \Sigma \over r^2 T^3 }
\end{equation}
 \begin{equation}
          = 1.02\times 10^{-3}
         \  r_{11}^{1/10}
         \  m_{{\rm BH}, \ 1}^{-7/10}
          \  \alpha_{-1}^{-1/5}
         \ {\dot m}_{-2}^{2/5}.
\end{equation}
  In summary,
  the standard model is
   marginally 
   self-consistent for  
   representative values in the outer disk
  $r_{11}\simeq {\dot m}_{-2} \simeq 1$.

  Note that for $ m_{{\rm BH}, \ 1} = r_{11}=1$, we have
  $\epsilon_{\rm acc}(r) \simeq  1.5\times 10^{-5}$
  and    ${\dot M}_E(r) \simeq  1.5 \times 10^{-3}\msunyr$.
       A value $ {\dot m}_{-2} \simeq 0.5$
  (determined from $L_{\rm II}$)
   at $r_{11}=1$
  would be locally super-Eddington only by a factor $\sim3$,
   but
      by larger factors at 
       smaller radii since ${\dot M}_E(r)\propto r$.
  As noted earlier, the true super-Eddington factors
  could be greater for the $r-$processed gas, 
  although for $T\simeq 10^7$ K the opacity is basically electron
scattering, which is actually
        suppressed due to the fact that the hydrogen
fraction $X=0$, and $\kappa_{\rm es} \propto (1+X)$.
  At smaller radii 
  the standard model 
 would become quite problematic
  due to 
$P_{\rm rad}/P_{\rm gas} >>1$
  leading to $h/r \ga 1$,
   and ${\dot M}/{\dot M}_E(r)  >>1$.
  Although a detailed 
  description of the disk at
smaller radii is not of  
primary relevance to this work,
    we briefly look at two 
  formalisms
   of potential interest.

\subsubsection{Super-Eddington Slim Disk}

For disks with accretion rates of order Eddington, 
 the  sizable radial 
gradients  
  demand the introduction of a 
   radial 
   advective
  energy transport term
(Abramowicz et al. 1988,
   Chen et al. 1995; 
   Kato, Fukue, \& Mineshige 1998).
  The cooling function
  is now written
\begin{equation}
Q^{-} = Q_{\rm rad}^{-} + Q_{\rm adv}^{-},
\end{equation}
where 
the advective transport
\begin{equation}
Q_{\rm adv}^{-} = \Sigma T v_r {ds \over dr},
\end{equation}
  and $s$ is the entropy per particle.
 This can be written as
\begin{equation}
Q_{\rm adv}^{-} = \Sigma  v_r {p\over \rho} \zeta,
\end{equation}
  where $\zeta$ is a logarithmic derivative
involving $s$ that is of order unity 
(Chen et al. 1995),  or
\begin{equation}
Q_{\rm adv}^{-} = {{\dot M} \over 2\pi r^2} c_s^2,
\end{equation}
where $c_s^2 = (3/2) GM/r$.
  The
 anticipated accretion rate $\sim 3 {\dot M}_{\rm E}(r)$
   at $r_{11}=1$
   places
  the accretion disk model
           on   the upper stable branch
 of slim disk solutions (Abramowicz et al. 1988, see their Fig. 4)
  where radiation pressure dominates
(Kato, Fukue, \& Mineshige 1998,
  see their section 10.1.1).
    Solving 
    the disk parameters
    for $P = P_{\rm rad}$
    gives
  \begin{equation}
\Sigma = {2 {\dot M} \over 9\pi 
                  r^2 \Omega_K
                  \alpha }
\end{equation} 
 \begin{equation}
   =  3.87 \times 10^4 \ {\rm g} \ {\rm cm}^{-2} 
         \  r_{11}^{-1/2}
         \  m_{{\rm BH}, \ 1}^{-1/2}
          \  \alpha_{-1}^{-1}
         \ {\dot m}_{-2},
\label{slim_sigma}
\end{equation}
 \begin{equation}
   T  = \left( 
   3 \Omega_K {\dot M} \over
   4\pi  a r \alpha
   \right)^{1/4}
\end{equation}
 \begin{equation}
       = 1.23 \times 10^6 \ {\rm K}
         \  r_{11}^{-5/8}
         \  {m_{{\rm BH}, \ 1}}^{1/8}
          \  \alpha_{-1}^{-1/4}
         \ {{\dot m}_{-2}}^{1/4},
\end{equation}
 \begin{equation}
 {h \over r}  = \sqrt{3\over 4} = 0.866,
\end{equation}
 \begin{equation}
\rho = 2.23 \times 10^{-7} \ {\rm g} \ {\rm cm}^{-3} 
         \  r_{11}^{-3/2}
         \  m_{{\rm BH}, \ 1}^{-1/2}
          \  \alpha_{-1}^{-1}
         \ {\dot m}_{-2},
\end{equation}
 \begin{equation}
 {P_{\rm rad}  \over P_{\rm gas}  } 
        =  4.05 \times 10^3
         \  r_{11}^{-3/8}
         \  m_{{\rm BH}, \ 1}^{7/8}
          \  \alpha_{-1}^{1/4}
         \ {\dot m}_{-2}^{-1/4},
\end{equation}
   and 
  \begin{equation}
  {Q_{\rm adv}^- \over Q_{\rm rad}^- }
   = {  \kappa_{\rm es} {\dot M}  
        \over  12  \pi c r }
         = { 1 \over 3 } { {\dot M} \over { {\dot M}_E(r)} }
        = 1.12
         \  r_{11}^{-1}
         \ {\dot m}_{-2}.
\label{Q_slim_ratio}
\end{equation}
  Thus the starting 
   assumptions $P_{\rm rad} > P_{\rm gas}$ 
   and $Q_{\rm adv}^-  >  Q_{\rm rad}^-$
    are validated.

\subsubsection{Marginally Eddington Disk with $P=P_{\rm rad}$}

In view of the super-Eddington accretion rates,
another possibility worth investigating
 is a disk in which strong mass outflow due to radiation pressure 
driven winds occurs
 as the local accretion rate begins to exceed ${\dot M}_E$.
Since the aspect ratio
\begin{equation}
 {h \over r} \simeq  { {\dot M}(r) \over {\dot M}_E(r) }
\end{equation}
(Shakura \& Sunyaev 1973),
   the maintenance of a  disk bordering on being
Eddington at all radii would enforce $h/r=1$.
     The disk would be non-steady state,
  \begin{equation}
{\dot M}(r) = {32\pi\over 3}  {c\over \kappa_{\rm es}}  r,
 \end{equation}
      and 
      strong winds would be generated
 at all radii. 
Solving for the disk
  variables with $P=P_{\rm rad}$ and  $h/r=1$  yields
 \begin{equation}
\Sigma = {16\over {3\alpha\Omega_K}}  {c\over {\kappa_{\rm es} r} } 
\end{equation}
 \begin{equation}
       = 6.94 \times 10^4 \ {\rm g} \ {\rm cm}^{-2} 
         \  r_{11}^{1/2}
         \  m_{{\rm BH}, \ 1}^{-1/2}
          \  \alpha_{-1}^{-1},
\label{windm_sigma}
\end{equation}
 \begin{equation}
  T  = \left( {3\over 2} 
     {r\Omega_K^2 \over a } 
       \Sigma
       \right)^{1/4} 
\end{equation}
 \begin{equation}
       = 1.16 \times 10^6 \ {\rm K}
         \  r_{11}^{-3/8}
         \  {m_{{\rm BH}, \ 1}}^{1/8}
          \  \alpha_{-1}^{-1/4},
\end{equation}
 \begin{equation}
 {h \over r}  = 1,
\end{equation}
 \begin{equation}
\rho = 3.47 \times 10^{-7} \ {\rm g} \ {\rm cm}^{-3} 
         \  r_{11}^{-1/2}
         \  m_{{\rm BH}, \ 1}^{-1/2}
          \  \alpha_{-1}^{-1},
\end{equation}
and
 \begin{equation}
 {P_{\rm rad} \over P_{\rm gas} } 
       = 2.20 \times 10^3  
         \  r_{11}^{-5/8}
         \  m_{{\rm BH}, \ 1}^{7/8}
         \  \alpha_{-1}^{1/4}.
\end{equation}
   Since there is no dependence on accretion rate, this state
   would persist, in theory, until ${\dot M}$ dropped below  ${\dot M}_E$
   at the outer edge.
   An annular region with constant ${\dot M}(r)$ would then 
   begin to propagate to smaller radii.
   This appears at first glance
   to raise the
   intriguing possibility
   of 
   applying this long period of constant accretion onto the central 
   engine 
   to explain the plateau phase of the light curve
   shown in Fig. 1 (segment II).
   If each annulus in the disk emits at roughly its local
   Eddington limit, an integration over the disk
   shows that the total
   accretion luminosity can exceed $L_E$
  by a factor $\sim\ln({\dot M}[r_{\rm outer}]/{\dot M}_E[r_{\rm inner}])$
  (Shakura \& Sunyaev 1973;
    Begelman, King, \& Pringle 2006)
   which would have a value $\sim10$ in the current application.
   However, the inferred GRB segment II luminosity
  $L_{\rm II} \simeq 3 \times 10^{43}$ erg s$^{-1}$$(f_{\rm beam}/3\times 10^{-3})$
   greatly exceeds even this enhanced Eddington value
  $\sim 10 \times L_E(10\msun) \sim 10^{40}$ erg s$^{-1}$,
   unless the beaming factor were orders
 of magnitude smaller than the value $3\times 10^{-3}$
 we take in this study.
     Such a tightly collimated flow 
   would not be expected from an Eddington outflow.

\subsection{Time Dependent Behavior of External Disks}

Pringle (1974, 1991) presented exact self-similar solutions
for  the time dependent behavior of disks with a free outer boundary
for the viscosity parameterization
\begin{equation}
\nu=\nu_0
       \left(\Sigma\over \Sigma_0\right)^m
       \left(r\over r_0\right)^n.
\end{equation}
 Of interest are the class of solutions
for which the disk total angular momentum is constant.

\subsubsection{Standard Model}

For the simplest case $-$  gas pressure and electron scattering opacity $-$
one may combine eqns. (2)-(4) to find
\begin{equation}
\nu= C r \Sigma^{2/3},
\end{equation}
where
\begin{equation}
 C = \alpha^{4/3} \left(k_B  \over \mu m_p\right)^{4/3} \left( \kappa_{\rm es} \over
    12acGM_{\rm BH} \right)^{1/3}.
\end{equation}
This is in the form considered by Pringle (1991) for
$m=2/3$ and $n=1$.
The solution is
\begin{equation}
{\Sigma(r,t)\over \Sigma_0}  
 = 
 \left( t\over t_0\right)^{-15/16} f\left[\left(
  r\over r_0\right)\left(t\over t_0\right)^{-3/8} \right],
\label{clg1990}
\end{equation}
where
\begin{equation}
 f(u) = \left(28\right)^{-3/2} u^{-3/5} \left(1-u^{7/5}\right)^{3/2}
\end{equation}
(Cannizzo, Lee, \& Goodman 1990),
  which can be verified by substitution into eqn. (\ref{evolution}).
The dimensional scalings $\Sigma_0$, $r_0$, and $t_0$
satisfy
\begin{equation}
 t_0 = r_0 C^{-1} \Sigma_0^{-2/3}.
\label{clg2}
\end{equation}
The solution given in eqn. (\ref{clg1990}) extends out to $u=1$,
so that the spreading of the outer edge
varies as $(r/r_0)=(t/t_0)^{3/8}$.
  The right hand side of  eqn. (\ref{clg1990}) multiplies
  the $\Sigma$ solution from eqn. (\ref{std_sigma}).
The total disk angular momentum is constant, and the disk mass
decreases as
\begin{equation}
 M_d(t) = (28)^{-3/2} {4\pi\over 7} 
r_0^2 \Sigma_0\left(t\over t_0\right)^{-3/16},
\end{equation}
  and therefore the 
available accretion luminosity
 $L_d(t) \propto {\dot M}_d \propto t^{-19/16}$.

 \subsubsection{Super-Eddington Slim Disk}
 
 In this model
 \begin{equation}
 \nu                               =  C_{\rm adv} r^{1/2},
 \end{equation}
 where
 \begin{equation}
          C_{\rm adv} = {9\over 4} \alpha  (GM_{\rm BH})^{1/2}.
 \end{equation}
 For
 $m=0$ and $n=1/2$
 the solution
  from  Pringle (1991; 
see his eqn. [3.2.8])  is
 \begin{equation}
 {\Sigma(r,t)\over \Sigma_0}  
  = 
 \left(r\over r_0\right)^{-1/2}
 \left(t\over t_0\right)^{-4/3}(1-k u^b)^a,
 \end{equation}
   where 
  $u=(r/r_0)^{1/2} (t/t_0)^{-1/3}$,
  $k=$
  $m(4m+4-2n)^{-1}$
  $ (5m+4-2n)^{-1}$
  $\propto m=0$,
  $a=1/m \rightarrow \infty$,
  and $b=3$.
 Euler's formula
 \begin{equation}
 \lim_{\mu\rightarrow\infty}
  \left(1 + {x\over\mu}\right)^{\mu}
 = e^x
 \end{equation}
 enables us to recast $\Sigma$ as
 \begin{equation}
 {\Sigma \over \Sigma_0}  
  = 
 \left(r\over r_0\right)^{-1/2}
 \left(t\over t_0\right)^{-4/3}
   \exp\left[-{1\over 9}
       \left(r\over r_0\right)^{3/2}
       \left(t\over t_0\right)^{-1}
        \right].
 \label{jeremy}
 \end{equation}
 Integration of
       $2\pi r dr \Sigma(r,t)$ 
   for the disk
 mass gives 
  \begin{equation}
 M_d(t) 
  = 
 12\pi r_0^2 \left(t \over t_0 \right)^{-1/3}
    \Sigma_0 
   \left(1 -
    \exp\left[-{1\over 9}
       \left(t\over t_0\right)^{-1}
        \right]
   \right).
 \end{equation}
  For $t<<t_0/9$,
     the $\exp$ term will go to zero.
 Hence one has $M_d \propto t^{-1/3}$ and therefore
   $L_d \propto t^{-4/3}$,
   in agreement  with Kumar et al. (2008a)
   for $s=0$.
  For late times  $t >> t_0/9$,
   expanding 
       $\exp(-x)$ as
   $1-x$ and combining the $t$ 
    dependencies 
    gives a steeper decay,
     $L_d \propto t^{-7/3}$.
  The dividing point $t_{\rm divide} = t_0/9$
    between  $L_d \propto t^{-4/3}$ and  $L_d \propto t^{-7/3}$
  can be evaluated by using the closure relation between 
  $C_{\rm adv}$,  $r_0$, and $t_0$, 
    yielding
 \begin{equation}
 t_{\rm divide}
  =  {16 \over 243}
   (\alpha \Omega_0)^{-1}  
       = 
    570 \ {\rm s} 
        \   r_{0, \ 11}^{3/2} 
        \   m_{{\rm BH}, \ 1}^{-1/2}
        \  \alpha_{-1}^{-1},
 \end{equation}
  where $\Omega_0=\Omega_K(r_0)$.

\subsubsection{Irradiation-dominated Disk}

At very late times corresponding to segment IV, which
 is not always seen,
    one needs also to consider irradiation-dominated disks.
For disks in which the locally defined irradiation temperature
$T_{\rm irr}=[L_{\rm acc}/(4\pi\sigma r^2)]^{1/4}$,
  where $L_{\rm acc}=\epsilon_{\rm acc} {\dot M} c^2$, 
exceeds the disk midplane temperature, the viscous heating
of the disk will be subordinate to irradiational heating.
One now finds
\footnote{Menou, Perna, \& Hernquist
(2001, see their Appendix A2)
   neglect the ${\dot M}$ (and hence $\Sigma$)
dependence of $\nu$, and therefore effectively
derive a similarity solution
  for a disk with constant irradiation.}
\begin{equation}
\nu= C_{\rm irr} r^{4/3} \Sigma^{1/3},
\end{equation}
where
\begin{equation}
 C_{\rm irr} = \alpha^{4/3} \left(k_B  \over \mu m_p\right)^{4/3} 
  \left( 4 \epsilon_{\rm acc}  c^2 \over
    27\sigma G^2M_{\rm BH}^2 \right)^{1/3}.
\end{equation}
This is in the form considered by Pringle (1991) for
$m=1/3$ and $n=4/3$.
The solution is
\begin{equation}
{\Sigma(r,t)\over \Sigma_0}  
 = 
 \left( t\over t_0\right)^{-5/3} f\left[\left(
  r\over r_0\right)^{1/2}\left(t\over t_0\right)^{-1/3} \right],
\end{equation}
where
\begin{equation}
 f(u) = (18)^{-3} u^{-2} \left(1-u^2\right)^3.
\end{equation}
The disk mass
 \begin{equation}
  M_d(t) =   (18)^{-3} {\pi\over 2} r_0^2 
   \Sigma_0\left(t\over t_0\right)^{-1/3},
 \end{equation}
so the disk luminosity $L_d \propto t^{-4/3}$.

\section{General Scenario}

We argue that the entire X-ray sequence spanning $>10^6$ s
can be understood in
terms of the external accretion disk passing through a
sequence of different physical states.
We propose that the jet
 becomes active almost immediately after
 creation of the BH
and the beginning
 of formation of the fall-back disk. The 
subsequent decrease in X-ray luminosity is then
due to the decreasing mass supply in the inner disk.
    The current paradigm  is that a
jet of material launched from the core of the 
progenitor decelerates as it 
encounters circumstellar material
(Sari et al. 1999; Frail et al. 2001).

In the following discussion of the four Zhang
et al. regimes, we adopt
GRB 060729
 as a specific example due to its
very long coverage both in terms
 of flux decay  and time (both span $\sim7$ decades
 $-$ see Grupe et al. 2007 for a detailed study). 
 The light curve
     (Grupe et al. 2007, see their Fig. 4;
 Willingale et al. 2007, see their Fig. 10)
shows a $0.3-10$ keV flux initially at $\sim 3 \times 10^{-7}$
erg cm$^{-2}$ s$^{-1}$ at
       $\sim 100$s, decreasing to $\sim 2\times 10^{-11}$
at  $\sim 10^3$s, then down slightly to $\sim 10^{-11}$ by  $\sim 5\times 10^4$s,
and finally to $\sim 3\times 10^{-14}$ at  $\sim 5\times 10^6$s.
The redshift $z=0.54$
gives, for the standard cosmology,
  a luminosity distance $d_L = 3.1$ Gpc, so that
$4\pi d_L^2 \simeq 1.15 \times 10^{57}$ cm$^{2}$.
   To
  take into account cosmological  time dilation
  the long term light curve depicted schematically
in Fig. 1 should also be 
   contracted along the $x-$axis by a factor $1+z=1.54$.
Adopting a geometric beaming factor
     $f_{\rm beam} \sim 3 \times 10^{-3}$,
   a beam    efficiency 
 $\epsilon_{\rm beam} = L_{\rm beam}/({\dot M}c^2) \sim 0.1$,
 and neglecting the bolometric correction,
the plateau flux 
    $F_{\rm XRT, \ II} \simeq 10^{-11}$ erg cm$^{-2}$ s$^{-1}$
translates into an
  isotropic
   luminosity $4\pi d_L^2 F_{\rm XRT, \ II}
  = 1.2\times 10^{46}$ erg s$^{-1}$,
      a beaming corrected
   luminosity $L_{\rm II} = 3.6\times 10^{43}$
 erg s$^{-1}$, 
   and a
  rest mass-energy flux
   in the disk ${\dot M} c^2 =
 L_{\rm II}/\epsilon_{\rm beam}
 = 3.6\times 10^{44}$  erg s$^{-1}$,
which would  not be directly observable
  presumably due to the advective
nature of the inner disk.
   This value is significantly  in excess
of the Eddington value $L_E \simeq 1.3 \times
10^{39}(M_{\rm BH}/10\msun)$ erg s$^{-1}$.
   It is also important to acknowledge
   that the very long duration of the plateau
  for GRB 060729, $\sim50,000$s,
  is not typical.
Dainotti et al. (2008) find a range between
about $10^2$s and $10^5$s in plateau duration, 
corrected   to the source frame.

Taking $f_{\rm beam} = 3 \times 10^{-3}$
    and     $ \epsilon_{\rm beam} = 0.1$
for entire evolution shown in Fig. 1 (which is probably unrealistic),
the amount of matter accreted 
during each segment
is
    $1.5\times 10^{-4}\msun$  for the prompt emission out to 100s,
    $5.5\times10^{-5}\msun$   for segment I,
    $8.5\times10^{-6}\msun$   for segment II,
    $2.4\times10^{-5}\msun$   for segment III,  and
    $1.7\times10^{-6}\msun$   for segment IV.
 It is important to note
  that these numbers represent the bookkeeping
    associated with the powering of the jet,
   which we are assuming produces the X-rays.
   Given the likelihood of significant mass depletion
   within the disk, e.g.,
  ${\dot M}(r) \propto r^s$, 
  with 
   $s$ in the range 0 to 1
    (Kumar et al. 2008a), 
   the actual total mass available
   from fall-back could
   be orders of magnitude larger, $\sim0.1-1\msun$.
Given the long term persistence  of
 the external disk,
the logarithmic rate of decay
for each of the four  segments identified by Zhang et al. (2006) 
must be relatable to different physical regimes
the disk passes through during its 
long period of mass depletion:

\subsection{Segment I: Fall-back Debris}

{$\alpha_{\rm I} \simeq  3$ for $10^2\la t\la10^3$ s}:
During this early time the inner disk is being dynamically
assembled out of fall-back material, and so the assumptions entering 
into the similarity solution for the evolution of the external disk
are not yet valid.
Kumar, Narayan, \& Johnson (2008a) 
considered the evolution of debris from the SN event that disrupts
the progenitor and creates the BH and subsequent GRB, and show 
analytically that the arrival of material
  onto the BH from the
collapse of the outer half of the progenitor core
 leads to 
${\dot M}\sim t^{-3}$.
    In their model the rapid infall of matter
 arises because of
the steep density profile of the outer core.
 During this early  period of extremely 
 high accretion rate
   $\ga  10^{10} L_{\rm E}$ 
  initially,
  the inner disk may contain regions
   where the cooling is
neutrino-dominated and 
   then advection-dominated.
   During the early, chaotic phase of fall-back and accretion,
  the high rate of accretion significantly depletes the amount of material
 surrounding the BH.

  It has been suggested that early in the evolution of the
  disk self-gravity may play a role (e.g., 
  Perna, Armitage, \& Zhang 2006).
Therefore
it may be possible to account for this early evolution
  purely within the
  confines of an accretion disk formalism,
  through considering
    self-gravitating disks (J. Pringle, 2008, private communication),
   although it is not clear a priori whether
such a steep decay law can be obtained.
If there is still substantial material at small radii
  left over from the immediate post-GRB accretion 
(MacFadyen \& Woosley 1999),
  it may be possible to model this decay time as a disk
  with self-gravity driving the
 viscosity.
     Detailed numerical calculations of
  self-gravitating disks support the contention that
the self-gravity torque can be efficient
  enough to maintain
  the disk at the edge of instability to self-gravity, 
   and that   the relevant physical length scales
   are small, of order the disk thickness $h$, 
   in which case
   a local viscosity formalism can be utilized
(e.g., Lodato \& Rice 2004).

\subsection{Segment II: Transient Plateau}

{$\alpha_{\rm II} \simeq  0.5$ for $10^3\la t\la10^4$ s}:
 The isotropic luminosity observed during the plateau $10^{46}$
erg s$^{-1}$ 
translates into a mass accretion rate 
 of $\sim 5 \times 10^{-3}  \msunyr$
  after applying the corrections given earlier.
 We argue that the plateau stage
occurs when a small amount of material
forms a torus at 
the radius of the progenitor, $\sim10^{11}$ cm.
  The
similarity solutions discussed 
earlier apply  
 to accretion disks that have had
time to adjust into a quasi-equilibrium
        profile which is close to steady-state.
   Therefore in practice 
 there will be  a period of adjustment
during which time
    the small amount of gas ($\ga 3\times 10^{-5}\msun$)
    that once was part of the 
envelope of the progenitor
  becomes stretched out due Keplerian shear
  into an 
   accretion disk.
 The time dependent calculations 
  of  Cannizzo et al. (1990)
  indicate a flat transient period in the light curve
   in response to an         annulus                  of material being
  suddenly
introduced near a central mass.
    An apparent problem with the light curves
  shown in Cannizzo et al. (1990, see their Fig. 3)
  is that the transient plateaus shown there have slightly
increasing fluxes, whereas the {\it Swift} XRT plateaus 
are flat or decaying. 
     The initial conditions 
 for their time dependent calculations
  were tori with 
  physically ad hoc
    Gaussian radial profiles,
  $\Sigma(r) \propto \exp[-(r-r_c)^2/(2w)^2]$,
    with $w/r_c = 0.5$.
   Therefore 
   time
  was required
    after the calculations started for matter 
  to spread 
  significantly
  to smaller radii,
     approach the standard disk
   profile $\Sigma(r)\propto r^{-3/4}$, 
   and raise the central accretion rate 
   onto the BH.
    For the present problem, however,
  it is more likely that
  one begins with significant amounts of gas at
 smaller  radii.
  Therefore the accretion rate 
  onto the BH would reach a higher level more quickly,
   and the light
    curves would be flat or decaying.
     Obviously,
    time-dependent
   accretion disk calculations
    for the present problem are desired,
   were a physically motivated initial $\Sigma(\vec{r},\vec{v})$ profile
   available from SN fall-back calculations.

Since our scenario begins
   to come into play at the beginning
 of segment II, the accreted masses quoted earlier for segments II through IV
must add up to
a minimum for our initial disk mass, i.e.,
                             about $3.5\times 10^{-5}\msun$.
  The relation between $\Sigma_0$, $r_0$, and 
$t_0$ given in eqn. (\ref{clg2}), in
combination with the numerical normalization
from  Cannizzo et al. (1990, see their Fig. 3),
can be used to form a scaling for this interval
\begin{equation}
 t_0 \simeq 5\times 10^4 \ {\rm s} 
         \  r_{0, \ 11}^{7/3}
          \  \alpha_{-1}^{-4/3}
         \  M_{d, \ -4.5}^{-2/3}
         \   m_{{\rm BH}, \ 1}^{1/3},
\label{plateau}
\end{equation}
where
$r_{0, \ 11}$
$=r_0/10^{11}$  cm,
   and
 $M_{d, \ -4.5} = M_{\rm disk}/10^{-4.5}$
$\msun$.
We argue 
that this plateau period
corresponds to
     Zhang et al's phase II.

\subsection{Segment III}

\subsubsection{Standard Disk}

{$\alpha_{\rm III}\simeq  1.2$ for $10^4  \la t \la  10^5$ s}:
This segment 
corresponds to the 
  standard model for the
  outer disk
   for which
the solution given earlier applies.
 Gas left over from the progenitor envelope
  at $\sim10^{11}$ cm
   has become smeared out
   and  evolved into the radial profile
  of an accretion disk, 
  and now the classical similarity solution for the
decay begins to apply.
 The disk mass varies as $t^{-3/16}$ so
the luminosity, which scales with $d M_d/dt$, varies as $t^{-19/16}\simeq t^{-1.2}$.
 For other opacity laws $\kappa=\kappa_0 \rho^a T^b$,  Cannizzo, Lee, \& Goodman
(1990) found $d\log L/d\log t = -(38+18a-4b)/(32+17a-2b)$.
This reduces to the electron scattering limit ($-19/16$)
for $a=b=0$. In the optically thick limit, the decay rate is relatively insensitive
to the opacity law. For instance, a 
 Kramer's law $a=1$, $b=-3.5$, gives a decay
 rate 
   $d\log L/d\log t = -5/4$.

\subsubsection{Super-Eddington Slim Disk/ADAF-like state}

For GRBs with high rates of accretion in the
post fall-back disk, conditions may be so extreme
that the standard model is not relevant.
Therefore the super-Eddington slim disk must be considered.
In section 2.4.2 
  we  present the time dependent solution 
for such disks, and show that for early times
   $d\log L/d\log t = -4/3$,
  which would probably be indistinguishable
  in the XRT data
  from the standard disk value
   $d\log L/d\log t = -19/16$.
Kumar, Narayan, \& Johnson (2008a;
 see their sections 3.1 and 3.2) take a
different approach in calculating the time
dependent evolution of the transient disk.
   Rather than solving the $\Sigma(r,t)$ evolution
equation   
  they prescribe a functional form ${\dot M}(r) \propto r^s$,
  and then solve for the total disk angular momentum, mass, and outer radius.
   The motivation for this  ${\dot M}(r)$ form
is that for ``ADAF-like''  conditions one expects strong mass outflow
  driven by a positive Bernoulli constant (Narayan \& Yi 1994, 1995).
   Thus $s=0$ would represent no mass loss, and $s=1$ would
correspond to a flow in which 90\% of the mass is lost
 from the disk for each inward decade in radius.
  This technique bypasses
  the need for a
detailed treatment of the
  $\Sigma(r,t)$ evolution equation
  and the kinematic viscosity $\nu$.
  It also ignores any potential transient phase
     through which the disk might have to evolve,
  and therefore assumes that the disk goes immediately
into its equilibrium state. 
    They derive a luminosity decay law
  $L_d(t)  \propto t^{-4(s+1)/3}$,
 which reduces to $t^{-4/3}$ for $s=0$.
   This agrees with our law for super-Eddington slim disks
  at 
early times.

\subsection{Segment IV}

\subsubsection{Irradiation Dominated Disk}

{$\alpha_{\rm IV}\simeq  2$ for $t \ga 10^6$ s}:
 Menou, Perna, \& Hernquist (2001) consider the evolution
 of supernova fall-back disks and find that irradiation
 becomes important at late times $t\ga10^6$s.
Since the radial decrease in
irradiation temperature is flatter than that for
disk temperature ($r^{-1/2}$ versus $r^{-3/4}$),
         as the disk mass decreases and the 
surface densities and temperatures in the outer
disk continue to drop, at some point the condition
$T_{\rm irr} > T_{\rm viscous}$
will be satisfied, and the standard model similarity solution
will be supplanted by the irradiation model solution.
Moreover, a strong wind will begin to deplete significantly 
the disk mass, at a rate that exceeds the local mass flow
(Begelman, McKee, \& Shields 1983,
 Begelman \&  McKee  1983,
Shields et al. 1986).
Therefore the decay rate 
will steepen 
  from the standard model
 value
$\alpha\simeq  1.2$.
 Begelman, McKee, \& Shields (1983, see their eqn. [4.2]),
derive a characteristic mass loss rate
due to a Compton heated wind
\begin{equation}
{\dot M}_{\rm ch} =  4.5\times 10^{-7} \ \msunyr \ \zeta L_{38} \left( L_{\rm cr}\over L \right)^{1/3},
\end{equation}
where $\zeta$ is a collection of factors of order unity, and 
$L_{\rm cr} \simeq 0.03 L_{\rm E}$.
Taking a  luminosity $L = \epsilon {\dot M}c^2$
gives
\begin{equation}
{\dot M}_{\rm ch} \simeq  2\times 10^{-3} \ \msunyr \ (\epsilon_{-1}{\dot m}_{-2})^{2/3}  m_{\rm BH, \ 1}^{1/3},
\end{equation}
 which will serve as a strong source of mass depletion
  in the outer disk.

The time-dependent decay
   solution for the wind-driven phase 
  presented in 
    section 2.4.3 is not complete because
  it does not include the effect of mass
decrease on $\Sigma(r,t)$. Also, the
   decay law, $\alpha_{\rm IV} = 4/3$,  is the same as the early-time
  solution for the super-Eddington slim disk,
  therefore could not account for the steepening
   decay for segment IV.
The generalized Kumar et al. (2008a) 
  decay solution
   in the presence of a disk wind, 
   $L_d(t)\propto t^{-4(1+s)/3}$,
  provides  more insight into the potential effect of a wind.
The fact that when a segment IV is observed it has a steeper
slope than segment III 
 suggests that if there are ADAF-like conditions
in the disk early on, mass loss is not significant since the observed
decay rate agrees with theoretical one if $s\simeq0$.
 Since the  
   ADAF-like state is supposed to be driven by a  positive
Bernoulli constant, as the disk mass drains and physical conditions
within the disk become less extreme, one would not expect for mass
loss to become stronger if the positive Bernoulli constant
   were the only force driving mass loss.
 However, 
 in the picture of 
 Menou, Perna, \& Hernquist (2001),
   in which irradiation becomes important at later times
   as a strong agent in driving a wind,
  a power law 
 ${\dot m}(r)
\propto r^s$ to characterize the
 variation of mass flow within the disk in the
presence of strong evaporation
may become relevant.
   The steeper decay law in segment IV would then indicate a
significant non-zero $s$ value: for example
a  nominal value $s\simeq0.5$ would be required to obtain
   $L_d(t)\propto t^{-2}$.

\subsubsection{Super-Eddington Slim Disk, late time solution}

In section 2.4.2 we found that for the time dependent
 super-Eddington slim disk, the disk mass 
for $t\ga 0.1  (\alpha\Omega_0)^{-1}$
  varies as  $M_d(t) \propto t^{-4/3}$, and therefore 
  $L_d(t) \propto t^{-7/3}$.
  This is close to the $\alpha_{\rm IV} \simeq 2$ from
Zhang et al. (2006) 
   and suggests another possibility for the late time
decay.  
In section 2.3.1 we presented the standard scalings for usual
accretion disk with $\nu\propto P_{\rm gas}$ and $\kappa=\kappa_{\rm es}$.
 A look at the various self-consistency tests showed that most of
them were at least marginally validated for conditions expected
in the outer disk during segment II. However, the range of 
   isotropic-equivalent
luminosities  for segment II (Dainotti et al. 2008) can range
up to $\sim2$ orders of magnitude greater than that adopted in section 2.3.1,
  which would  place an uncomfortable strain on the standard model.
   For GRBs with outer disks at such extreme conditions that
  the standard model does not apply, the super-Eddington slim
disk maybe a better description.   
    The range in $L_{\rm II}$ values shown in Dainotti et al. (2008)
    appear to show  an intrinsic scatter, therefore there may be
GRBs for which the standard model does apply at late times (low $L_{\rm II}$),
and others for which the slim disk model is required (high $L_{\rm II}$).
    One prediction of this idea would be that intrinsically brighter
bursts would be more likely to exhibit a segment IV, with its steeper
decay.  However, 
  for the fainter bursts one would not be able to follow
the late-time decay to as low a flux level, and therefore it might be
difficult to make a definitive observational 
  statement about the presence of a
segment IV.

\subsection{Mass supply: Viscous accretion vs. Fall-back}

 Kumar et al. (2008b) present and discuss the criteria for
   determining whether viscous accretion or fall-back will be the
primary supply mechanism feeding mass into the central engine.
 Both processes
will influence the light curve, 
  but the one with the longer time scale is likely to have
the dominant effect. 
   After the SN has occurred,
    the fall-back time for an element of gas in the progenitor
  envelope to fall from radius $r$ to the BH
  is roughly the free-fall time $t_{\rm fb}\sim2(r^3/GM_{\rm BH})^{1/2}$.
(This ignores any residual pressure support during the collapse,
  and other effects such as a wind from the accretion disk
  and a shock originating near the BH during its formation.)
Inverting this expression gives the radius in the star 
 expected to be accreting at a given time $t$ after the SN occurs, 
$r_{10} \simeq 1.5 t_2^{2/3}  m_{\rm BH, \ 1}^{1/3}$,
   where $t_2=t/100$ s.
  The specific angular momentum 
$j_{*,18} = 3.6  m_{\rm BH, \ 1}^{1/2}  r_{10}^{1/2} f_{\Omega}(r)$,
  where $j_{*,18}$ is  
  in units of $10^{18}$ cm$^2$ s$^{-1}$, and 
  $f_{\Omega}(r)$ is the local angular velocity of a gas at radius
   $r$ in the progenitor, in terms of the Keplerian value $\Omega_K$.  
  After the SN explosion, gas which ends up being trapped in the potential well
of the BH will fall inward to the point where it hits its angular momentum
barrier, determined  by $f_{\Omega}(r)$.
   This viscous accretion time scale 
  $t_{\rm acc} \simeq (r/h)^2 (\alpha\Omega_K[r_{\rm circ}])^{-1}$
$\approx 2  \alpha^{-1}{\Omega_K}^{-1}(r_{\rm circ})$ (since $h/r\simeq 1$),
where $r_{\rm circ}$ is the circularization radius $\sim j_*^2/(GM_{\rm BH})$.
   This line of reasoning leads to eqn. [4] of Kumar et al. (2008b),
 \begin{equation}
 t_{\rm acc}/t_{\rm fb} \simeq 10 \alpha_{-1}^{-1} {{f_{\Omega}}^3(r)}.
 \label{knjb}
 \end{equation}
    Kumar et al. (2008b) note that this relation implies the accretion time 
will be shorter than the fall-back time,  and therefore subordinate
as a mass supply mechanism, for $f_{\Omega}(r) \la 0.4$.
  They find that the steep decay segment (I) would be possible only if
the disk can adjust rapidly to a decrease in ${\dot M}_{\rm fb}$,
  and use this consideration to place an upper limit on
the rotation speed in the core (on the assumption that segment I is
due to fall-back of the progenitor core). 
   Their approach is strengthened by our
finding that all plausible disk states give 
 decay rates with power laws $\la 2$,
too shallow to explain segment I.

At early times before the disk has had a significant opportunity to spread,
the fundamental physical considerations  presented by  Kumar et al. (2008b) 
  seem well-motivated.
  At later times, however, the increasing viscous time scale associated
with the outer disk edge, and the maintenance of a causal coupling to
smaller radii, via angular momentum transport, begin to introduce a
significant correction factor that must be taken into account in 
eqn. (\ref{knjb}). It is important to note that
        implicit in the derivation of eqn. (\ref{knjb})
is that the radius appearing in the fall-back time and the
viscous time are the same. However, the spread of the outer disk
edge introduces a strong nonlocality:  
   instead of taking 
      $t_{\rm acc}\propto\alpha^{-1} {\Omega_K}^{-1}(r_{\rm circ})$ 
  one should more properly
               consider $\alpha^{-1} {\Omega_K}^{-1}(r_{\rm disk, \ outer})$.
   Since 
      $t_{\rm acc}$ varies as $r^{3/2}$,
this introduces  a correction factor $(r_{\rm disk, \  outer}/r_{\rm circ})^{3/2}$
into eqn. (\ref{knjb}).
  Taking a fiducial spreading rate $(r/r_0)=(t/t_0)^{3/8}$ as 
in the standard solution, and adopting typical numbers
 for times separating the different segments, $t_{\rm  I/II} \simeq         10^3$s,
                                              $t_{\rm II/III}\simeq 5\times 10^4$s,
                                         and  $t_{\rm III/IV}\simeq 3\times 10^6$s,
this implies an increase over the locally defined (i.e., at $r_{\rm circ}$) accretion time
scale by a factor $\sim 8$ at $t_{\rm II/III}$
           and $\sim 90$   at $t_{\rm III/IV}$,
   respectively (stemming from increases of $r_{\rm disk, \ outer}$
  by a factor 4 from $t_{\rm I/II}$ to $t_{\rm II/III}$,
  and
  by a factor 20 from $t_{\rm I/II}$ to $t_{\rm III/IV}$).
  The faster rate of expansion
   calculated by  Kumar et al. (2008a), 
  $(r/r_0)=(t/t_0)^{2/3}$, 
    would give larger correction factors.
   Also, these factors are actually lower limits because
  they only take into account the expansion of the
   disk relative to $t_{\rm I/II}$.
   The disk expansion enhances the  effect of the accretion
   supply with respect to the mass supply by 
  prolonging $t_{\rm acc}$,  and also
facilitates an understanding of why the steepest decay (segment I)
      must occur at
the earliest time possible,
        before the significant outward movement 
        in $r_{\rm disk, \ outer}$ has enforced $t_{\rm acc} >> t_{\rm fb}$.

     Kumar et al. (2008b)
   present examples of 4 {\it Swift} GRBs for which $\alpha_{\rm III}\simeq4-6$,
much steeper than in the standard 
       Zhang et al. (2006) picture  $\alpha_{\rm III}\simeq1.2$.
  For these rare bursts the 
    Kumar et al. fall-back scenario may well provide a better
description than our viscous accretion disk scenario.

\section{Discussion}

The overarching question behind this study
is the following:  Is the very long term ($t\ga 10^6$s)
decay seen in the X-ray flux from GRBs due to 
   the deceleration of a relativistic jet,  or to
  a decreasing mass supply onto the central engine?
  We argue in favor of the latter.
We have presented a general scenario for understanding the long term
evolution,
  in particular the 
four  power law decay segments found through {\it Swift} XRT observations.
   The basic idea relies
   on the fact that the ever-expanding
outer edge maintains a causal connection
      through the intervening radii
all the way to the inner disk, given the coupling of the viscosity mechanism
to the local shear in the disk.
   In more basic terms, the mass of the accretion disk decreases
   as a power law in time, resulting in a decrease in
mass accretion rate at the inner edge.

The overall evolution is as follows:

(I) The initial steep decay phase arises from the dynamical fall-back
of $\sim0.1\msun$ of material into the vicinity of the BH
(Kumar, Narayan, \& Johnson 2008ab).
 Kumar et al. 2008b
    show by comparing the viscous
  and fall-back times in the early stages of fall-back that
  $t_{\rm acc}<<t_{\rm fb}$. The disk 
       therefore responds quickly
   to the steeply decreasing, externally 
imposed mass supply from the progenitor core.
    This scenario is only possible at early times before
 expansion of the outer accretion disk causes 
  $t_{\rm acc} > t_{\rm fb}$.
 An alternative but perhaps less likely 
   possibility for the steep decay  involves 
completely different physics 
(Pe'er et al.\footnote{
           Their explanation, which involves a 
hot cocoon with modest Lorentz factor and
a nearly thermal spectrum,
    may  be refuted
   by {\it Swift} observations 
 showing  that the GRB prompt emission joins smoothly
 to segment I $-$ an unlikely coincidence
   given that the hot cocoon represents
  an independent physical component.}
        2006).

(II) 
      The initial establishment of the disk leads to a 
brief transient phase during which calculations
indicate a flat light curve.
  This corresponds to the viscous smearing
 and spreading of gas
 left over from the progenitor
envelope
  into an accretion disk (Cannizzo et al. 1990), 
which we take to lie at roughly the progenitor
  radius $\sim10^{11}$ cm.
   The dynamical time scale 
  there
        $\Omega_K^{-1}\sim 10^3$s
  is toward the lower range of observed plateau durations 
  (Dainotti et al. 2008, se their Figs. 1 and 2),
  and
 our scaling for the transient period
   following the arrival of material
 to establish the accretion disk
  is in line with the duration
of the plateau phase seen in some of the
   XRT light curves.

   The plateau  phase has been ascribed by previous workers
to continued activity of the central engine,  or to the accretion
disk having low value of the
  Shakura-Sunyaev
   $\alpha$ parameter
(e.g., Kumar, Narayan, \& Johnson 2008a;
    although in their final analysis they
    downgrade this possibility).
    Given 
  the apparent robustness of the non-linear saturation limit of the viscosity
parameter in the MRI process
(Beckwith, Hawley, \& Krolik 2008),  
    however, 
   we also view the small  $\alpha$  explanation
  as unlikely. 
   Kumar, Narayan, \& Johnson (2008a)
  view as a more plausible explanation 
continued fall-back onto the central BH.
   Kumar, Narayan, \& Johnson (2008b; see their Fig. 2)
 present conditions for the core/envelope density structure
 of the progenitor that would be required to obtain the plateau
(as well as the steeper segments - I and III).
  In the previous section we showed how the spreading of the outer
disk enhances the effective viscous accretion time, relative to
the fall-back time, and ensures that viscous accretion will
   become increasingly dominant as the outer disk expands significantly
beyond  what had been the radius of the progenitor.

(III) After the transient period, 
   the outer disk becomes describable
   using the standard formalism
    ($t^{-1.2}$ decay)
     or an advective formalism
     ($t^{-1.3}$ decay).
 The continued high rate of accretion 
substantially depletes the 
mass of the disk.
  The inner disk is still
  not only substantially super-Eddington (by a factor $\sim10^5 - 10^7$)
   but also potentially degenerate
   due to the very high densities
   (e.g., 
   Narayan, Piran, \& Kumar 2001,
  Kohri \& Mineshige 2002,
  Kohri, Narayan, \& Piran 2005,
   Chen \& Beloborodov  2007,
   Ohsuga \& Mineshige 2007).

(IV) At some point the outer disk may become 
not only radiation dominated, but also subject
   to a substantial Compton driven wind,
      leading to an
increase in the decay rate.   
   If the rate of mass flow within the body
of the disk can be characterized ${\dot M}(r) \propto r^s$
as in Kumar et al. (2008a), then their
  decay solution $L_d(t)\propto t^{-4(1+s)/3}$ would
provide a good fit for $s\approx  0.5$ 
   regardless of whether the disk was standard or advective.
    Alternatively, our late time super-Eddington slim-disk
solution  $L_d(t)\propto t^{-7/3}$ may apply at this time.

Depending on details such as the initial  mass of the disk, 
        states (II) and (IV)  may or may not
  become manifest in the long term light curve.

The current paradigm for
the long term GRB light curves
attributes changes seen in the decay properties to variations (i.e., a decrease)
in the bulk Lorentz flow of a jet of material expelled from the central engine
as it collides with and interacts with circumstellar material.
  We argue that a strong jet is established almost immediately after the
GRB  prompt emission, and the subsequent decay properties of the light curve
are set by a dwindling mass supply  
due to the long term decrease in $\Sigma$ at all radii.  
   The decay law obtained by Kumar et al. (2008a) in which
 $L_X\propto t^{-4(s+1)/3}$, where the mass loss rate from the disk
varies as $r^s$, allows for a range of decay between $t^{-4/3}$ and  $t^{-8/3}$.
   The steeper decay, as an explanation for segment III, seems
  to be excluded by the data (e.g., Ghisellini et al. 2009, see their Fig. 16).
  If the slim disk/ADAF picture proves to be a valid description of the 
   outer disk, the observed segment III decay rate would appear to 
argue against a strongly varying mass loss rate with radius.
    Also, the potential bimodality 
   in decay rate with time found in section 
 2.4.2 for the slim disk state,
    $t^{-4/3}$ for early times and  $t^{-7/3}$ for late times,
    might play a role in the III/IV transition.
  The dividing point lies at 
 $\sim10^3$s $r_{0, \ 11}^{3/2}  m_{{\rm BH}, \ 1}^{-1/2} \alpha_{-1}^{-1}$,
  so for this to work the disk would have to have expanded to 
$\ga10^{12}$ cm to account for the long time scale $\sim10^4-10^5$s
  of the III/IV transition.
  Alternatively, the onset of a strong wind 
at late times,
      driven by irradiation, 
   may  force $s$ to increase from $0$ to $\sim0.5$, 
   thereby steepening the decay in segment IV.

  In addition to our alternative explanation for the plateau seen in some
 of the XRT light curves, there is also an issue raised by the high-latitude emission
model (Kumar \& Panaitescu 2000). In this model, after the mechanism
responsible for producing X-ray emission ceases,
an observer along the central jet axis sees continued
emission
as photons from increasingly off-axis lines of sight continue to arrive.
For uniform surface brightness, an intrinsic spectrum
$\nu^{-\beta}$ produces a power law decay in time
$t^{-\alpha}$, where $\alpha=\beta+2$.
This prediction was tested by O'Brien et al. (2006, see their Fig. 4)
  for the initial decay segment
(Zhang et al's I) using a large sample of {\it Swift} GRBs. 
Instead of 
   $\alpha\simeq\beta+2$, they found
a scatterplot
    in $\alpha$ versus $\beta$.  
  Thus
  the standard model also seems problematic for decay segment I
as regards the temporal/spectral evolution,
  although more recent work on the high-latitude ``curvature effect''
   has been promising (Zhang et al. 2007, 2009).

\section{Implications} 

  Eqn. (1) of Sari, Piran, \& Halpern (1999)
  has proven of great use to observers insofar as
 it relates observable quantities, namely 
 jet break time, redshift, and GRB fluence, to a beaming 
angle  (assuming a density 
  of the circumstellar medium).
   The
    outline of 
   the  theory we present
  is not yet sufficiently developed to build a 
  complex
   formalism.
   Simple statements can still
be made, however. 
   Our eqn. (\ref{plateau}) gives an approximate scaling for
  the time of the segment II/III transition. 
  If the luminosity of the plateau scales roughly
with the mass accumulated during fall-back that
supplies the transient disk associated with the plateau
 (segment II) in our model, then eqn. (\ref{plateau})
   predicts an inverse scaling
  between $L_{X, {\rm II}}$ and $t_{\rm II, \ III}$,
  namely
          $L_{X, {\rm II}} \propto  t_{\rm II, \ III}^{-3/2}$.
   (This ignores the $r_0$ scaling, however.)
   An  inverse correlation is observed in the data,
   but appears less steep 
         ($L_{X, {\rm II}} \propto  t_{\rm II, \ III}^{-0.8}$)
    than our prediction 
 (Dainotti, Cardone, \&  Capozziello 2008;
   see their Figs. 1 and 2).

The Kumar et al. (2008a)
  decay law with mass loss from the disk, 
   $L_d(t)\propto t^{-4(1+s)/3}$,
  is consistent with our early time decay law
  for the super-Eddington slim disk, and with the
  observed value $\alpha_{\rm III} \simeq 1.2$,  for $s=0$.
   It is also  close to the decay law for standard
disks,  $L_d(t)\propto t^{-19/16}$.
   This appears to indicate that for very high ${\dot M}$ disks
  where the standard model is not applicable during segments
II and III, if an ADAF-like state is correct, then mass loss
  must be minimal (i.e., $s\simeq0$, or ${\dot M}(r)\simeq$ constant).
   The steepening then associated with segment IV could
either be due to the onset of significant evaporation, e.g., $s\simeq0.5$
would be sufficient, or the late time super-Eddington slim disk law
    $L_d(t)\propto t^{-7/3}$.

If our model has some merit, then a revision in
current thinking would be needed.
   First, there may not be a standard GRB energy reservoir 
      as in Frail et al. (2001).
      Also,
      for the jet to respond quickly
      to whatever conditions prevail at the central engine
(i.e., accretion rate),  it may need to be a light-weight
jet  with a low degree of baryon loading. 
    It may be
    a  Poynting-flux dominated jet as in the model
  of Lyutikov \& Blandford (2003).
 The deceleration experienced by such a jet would tend to be less,
  for a given energy flux.
   The previous estimates of 
the circumstellar density field $n(r)$ would need to be revised.
 Conversely, the jet may be baryon-loaded during the
steep decay segment (segment I) which appears to be too steep to be covered
within an accretion disk formalism.
   The spectral evolution observed
  during this phase may be difficult to account for
   with a purely EM jet (O'Brien et al. 2006; Pe'er et al. 2006).
  There may also be some baryon loading during the rest
of the decay, but the density of the circumstellar medium
would need to be less than previously thought, so that deceleration
would be  minimal.

  In this work we have implicitly assumed that the X-ray light
curve represents the bulk of the radiated jet energy.
  Therefore our model runs into the same  difficulty
  as the standard model in accounting for why the breaks seen
 at other wavelengths do not follow the X-ray.
  The physical emission  mechanisms
   must differ across different wavebands,
   which could also potentially mitigate one of the
  criticisms of the standard model.
   One idea currently being discussed is that, 
   within a given GRB
 there
  can co-exist jets with different opening angles
  and different physical conditions,
   each of which can contribute
   preferentially to emission in different wavebands.
     If the  opening angles  
 are considerably different, 
    this could lead to different break times
   (e.g., Racusin et al., 2008).
      Our simple model is not 
   yet developed enough to make any predictions
    on the relative rates of decay
  at different wavelengths.

  There are other important issues to consider, 
  such as the formation and stability of an accretion
  disk in the middle of a supernova.
   Enough material would have to be evacuated from the
  vicinity
  of the disk near the BH
   at early times to permit the disk to survive
   the expected shocks and radiation of the explosion.
   Also, if the BH acquires a kick velocity $v_{\rm kick}$ during the initial 
explosion, this velocity would have to be small enough so that the accretion
   disk was not left behind.
 (A simple consideration of the binding 
  energy shows that
   matter in the disk with  
   $v_{\rm orbit} = 1155$ km s$^{-1}$ 
   $m_{{\rm BH}, \ 1}^{1/2}$
   $r_{11}^{-1/2} > v_{\rm kick}$,
 where $ v_{\rm kick} \simeq 100-300$
      km s$^{-1}$,
       should remain bound to the BH.)
   Finally, the jet issuing from the BH 
   would have to be powerful enough to punch
through the veil of material comprising the ejecta, and to maintain
  an open channel for $\ga10^6$s.
    If the jet is primarily electromagnetic as opposed to baryon-loaded,
   the open EM field lines
                    may help in maintaining a clear path for the jet beam.

Recently various workers have proposed a simple way for
obtaining the flat decay associated with segment II.
  Implicit in our work and that of  others such as 
Kumar et al. (2008ab) is that the start of the prompt emission
   is the relevant start time  
     for all the other emission
    $-$  X-ray,
optical, and radio.
   It is possible that the physical mechanism
   responsible for producing the X-rays
   actually commences $\sim10^4 - 10^5$s before the prompt emission, so
that the flat appearance of segment II is merely due to having 
an incorrect start time (Yamazaki 2009; Liang et al. 2009).
  Yamazaki (2009)
   plots four examples of GRBs for which a shifting of the start
time for the X-ray decay produces a good fit for the plateau,
   after having optimized the decay for segment III.
    One can however find other examples of GRBs for which there 
    is a gradual decay in segment II 
    that is not well-fit
by simply shifting the start time, therefore this explanation may not
work for all GRBs. Also, it remains unclear as to 
  how the X-ray emission
  could begin significantly before 
the GRB, which  presumably marks
the moment of explosion of the progenitor and formation of the BH.

\section{Conclusion} 

We present a new paradigm
for understanding the long-term
behavior of GRBs, as inferred primarily
through {\it Swift} XRT observations.
We argue that the four  power law segments into which
the long term XRT light curves can be divided 
can be explained by the time dependent evolution
of the external accretion disk.
The initial, steep period of decay 
   happens
 before the external accretion disk has been
established and therefore is beyond our scope
   (although a self-gravitating disk remains a possibility).
   Simple estimates of the accretion
  and fall-back time scales 
     $t_{\rm acc}$ and  $t_{\rm fb}$ 
              indicate that
  fall-back will be dominant at the earliest times.
  Therefore the fact that segment I has the steepest slope,
  too steep to be accounted for by the disk models we study,
  strengthens the picture advocated in Kumar et al. (2008b)
  in which the direct fall-back of the progenitor core
      leads to segment I.  After the initial evolution,
  expansion of the outer disk edge enforces  
 $t_{\rm acc} > t_{\rm fb}$ so that accretion is dominant.
   We ascribe the subsequent flattening in the GRB
light curves to a period of viscous re-adjustment in 
the accretion disk formed out of the 
  gas elements from the 
progenitor envelope that  remain  bound to the BH
  and end up at their
   specific angular momentum radii 
   after the hypernova explosion.
    The following steepening corresponds to the similarity 
solution for the decay of a standard disk 
   or advective disk
   at the largest radii,
    and the further slight steepening 
  at later times could be due
to the additional mass depletion from a strong wind,
  or in some cases to the late time evolution of a 
  super-Eddington slim disk.

\acknowledgements

We thank Kris Beckwith,
   Juhan Frank,
  Chris Fryer,
  Pranab Ghosh, 
  Josh Grindlay,
  Andrew King,
  Shin Mineshige, 
  Jim Pringle,
  Judy Racusin,
  Brad Schaefer,
  and Jan Staff
   for useful discussions.
We thank Jeremy Goodman for  
help with eqn. (\ref{jeremy}).
   We thank the referee for a careful reading of the paper
and very useful comments.

\def\mnras{MNRAS}
\def\apj{ApJ}
\def\apjs{ApJS}
\def\apjl{ApJL}
\def\aj{AJ}
\def\araa{ARA\&A}
\def\aap{A\&A}


\begin{thebibliography}{75}
\expandafter\ifx\csname natexlab\endcsname\relax\def\natexlab#1{#1}\fi

\bibitem[]{}
Abramowicz, M.~A., Czerny, B., Lasota, J.-P., \& Szuszkiewicz, E. 1988, ApJ, 332, 646

\bibitem[]{}
Balbus, S.~A., \& Hawley, J.~F. 1998, Rev Mod Phys, 70, 1

\bibitem[]{}
Balbus, S.~A., \& Papaloizou, J.~C.~B. 1999, ApJ, 521, 650

\bibitem[]{}
Beckwith, K., Hawley, J.~F., \& Krolik, J.~H. 2008, ApJ, 678, 1180

\bibitem[]{}
Begelman, M.~C., King, A.~R., \& Pringle, J.~E. 2006, MNRAS, 370, 399

\bibitem[]{}
Begelman, M.~C., \& McKee, C.~F. 1983, ApJ, 271, 89

\bibitem[]{}
Begelman, M.~C., McKee, C.~F., \& Shields, G.~A. 1983, ApJ, 271, 70

\bibitem[]{}
Blandford, R.~D., \& Znajek, R.~L. 1977, MNRAS, 179, 433

\bibitem[]{}
Blustin, A.~J., et al. 2006, ApJ, 637, 901

\bibitem[]{}
Burrows, D.~N., et al. 2005, Science, 309, 1833

\bibitem[]{}
Cannizzo, J.~K. 1997, ApJ, 482, 178

\bibitem[]{}
Cannizzo, J.~K., Lee, H.~M., \& Goodman, J. 1990, ApJ, 351, 38

\bibitem[]{}
Cannizzo, J.~K., \& Reiff, C.~M. 1992, ApJ, 385, 87

\bibitem[]{}
Chen, W.-X., \& Beloborodov, A.~M. 2007, ApJ, 657, 383

\bibitem[]{}
Chen, X., Abramowicz, M.~A., Lasota, J.-P., Narayan, R., \& Yi, I. 1995, ApJ, 443, L61

\bibitem[]{}
Curran, P.~A., et al. 2007, MNRAS, 381, L65

\bibitem[]{}
Dai, X., Halpern, J.~P., Morgan, N.~D., Armstrong, E.,
  Mirabal, N., Haislip, J.~B.,
  Reichart, D.~E., \& Stanek, K.~Z.
      2007, ApJ, 658, 509

\bibitem[]{}
Dainotti, M.~G.,  Cardone, V.~F., 
        \& Capozziello, S. 2008, MNRAS, 391, L79

\bibitem[]{}
Done, C., Gierli\'nski, M., \& Kubota, A. 2007,
 Astr \&  Astrophys Rev, 15, 1

\bibitem[]{}
Frail, D.~A., et al. 2001, ApJ, 562, L55

\bibitem[]{}
Fryer, C.~L.,  Colgate, S.~A., \& Pinto, P.~A. 
   1999, ApJ, 511, 885

\bibitem[]{}
Gehrels, N., et al. 2004, ApJ, 611, 1005

\bibitem[]{}
Gehrels, N., Ramirez-Ruiz, E., \& Fox, D.~B. 2009, ARAA, in press 

\bibitem[]{}
Ghisellini, G., Nardini, M., Ghirlanda, G., \& Celotti, A. 2009, MNRAS, 393, 253

\bibitem[]{}
Ghosh, P., \& Abramowicz, M.~A. 1997, MNRAS, 292, 887

\bibitem[]{}
Grupe, D., et al. 2007, ApJ, 662, 443

\bibitem[]{}
Hirose, S., Krolik, J.~H., \& Blaes, O. 2009, ApJ, 691, 16

\bibitem[]{}
Kato, S., Fukue, J., \& Mineshige, S. 1998,
   Black-Hole Accretion Disks
   (Kyoto Univ. Press; Kyoto)

\bibitem[]{}
Kawanaka, N., \& Mineshige, S. 2007, ApJ, 662, 1156

\bibitem[]{}
Kocevski, D., \& Butler, N. 2008, ApJ, 680, 531 

\bibitem[]{}
Kohri, K.,    \& Mineshige, S. 2002, ApJ, 577, 311

\bibitem[]{}
Kohri, K., Narayan, R.,  \& Piran, T. 2005, ApJ, 629, 341

\bibitem[]{}
Krolik, J.~H., Hawley, J.~F., \& Hirose, S. 2007, Rev. Mex. A. A., 27, 1

\bibitem[]{}
Kumar, P., Narayan, R., \& Johnson, J.~L. 2008a,  MNRAS, 388, 1729

\bibitem[]{}
Kumar, P., Narayan, R., \& Johnson, J.~L. 2008b, Science, 321, 376

\bibitem[]{}
Kumar, P., \& Panaitescu, A. 2000, ApJ, 541, L51

\bibitem[]{}
Lazzati, D., Perna, R., \& Begelman, M.~C. 2008, MNRAS, 388, L15

\bibitem[]{}
Liang, E.-W., L\"u, H.-J., Zhang, B.-B., \& Zhang, B.
              2009, astro-ph/0902.3504

\bibitem[]{}
Liang, E.-W., Racusin, J.~L., Zhang, B., Zhang, B.-B.,
           \& Burrows, D.~N.
              2008, ApJ, 675, 528

\bibitem[]{}
Lightman, A.~P., \& Eardley, D.~M.             1974, ApJ, 187, L1

\bibitem[]{}
Livio, M., Ogilvie, G.~I., \& Pringle, J.~E.   1999, ApJ, 512, 100

\bibitem[]{}
Lodato, G., \& Rice, W.~K.~M. 2004, MNRAS, 351, 630

\bibitem[]{}
Lyutikov, M., \& Blandford, R.~D. 2003, astro-ph/0312347

\bibitem[]{}
MacFadyen, A.~I. \& Woosley, S.~E. 1999, ApJ, 524, 262

\bibitem[]{}
McKinney, J.~C., \& Narayan, R. 2007a, MNRAS, 375, 513

\bibitem[]{}
McKinney, J.~C., \& Narayan, R. 2007b, MNRAS, 375, 531

\bibitem[]{}
Menou, K., Perna, R., \& Hernquist, L. 2001, ApJ, 559, 1032

\bibitem[]{}
Nagataki, S., Takahashi, R., Mizuta, A., \& Takiwaki, T. 
  2007, ApJ, 659, 512

\bibitem[]{}
Narayan, R., Piran, T., \& Kumar, P. 2001, ApJ, 557, 949

\bibitem[]{}
Narayan, R., Yi, I. 1994, ApJ, 428, L13

\bibitem[]{}
Narayan, R., Yi, I. 1995, ApJ, 444, 231

\bibitem[]{}
Nousek, J.~A., et al. 2006, ApJ, 642, 389

\bibitem[]{}
Oates, S.~R., et. al. 2007, MNRAS, 380, 270

\bibitem[]{}
O'Brien, P.~T., et al. 2006, ApJ, 647, 1213

\bibitem[]{}
Ohsuga, K., \& Mineshige, S. 2007, ApJ, 670, 1283

\bibitem[]{}
Pe'er, A.,  M\'esz\'aros, P., \& Rees, M.~J. 2006, ApJ, 652, 482

\bibitem[]{}
Perna, R., Armitage, P.~J.,   \& Zhang, B. 2006, ApJ, 636, L29

\bibitem[]{}
Pringle, J.~E. 1974, PhD thesis, Univ. of Cambridge

\bibitem[]{}
Pringle, J.~E. 1991, MNRAS, 248, 754

\bibitem[]{}
Racusin, J.~L., et al. 2008, Nature, 455, 183

\bibitem[]{}
Rhoads, J.~E. 1997, ApJ, 487,  L1

\bibitem[]{}
Rhoads, J.~E. 1999, ApJ, 525, 737

\bibitem[]{}
Sakamoto, T., et al. 2008, ApJS, 175, 179

\bibitem[]{}
Sari, R., Piran, T., \& Halpern, J.~P. 1999, ApJ, 519, L17

\bibitem[]{}
Shakura, N.~I., \& Sunyaev, R.~A. 1973, A\&A, 24, 337

\bibitem[]{}
Shields, G.~A., McKee, C.~F.,
 Lin, D.~N.~C., \& Begelman, M.~C. 1986, ApJ, 306, 90

\bibitem[]{}
Smak, J. 1984, Acta Astr., 34, 161

\bibitem[]{}
Stanek, K.~Z., et al. 2007, ApJ, 654, L21

\bibitem[]{}
Willingale, R., et al. 2007, ApJ, 662, 1093

\bibitem[]{}
Woosley, S.~E.  1993, ApJ, 405, 273

\bibitem[]{}
Woosley, S.~E., \& Bloom, J.~S. 2006, ARAA, 44, 507

\bibitem[]{}
Yamazaki, R. 2009, ApJ, 690, L118

\bibitem[]{}
Yuan, F., Ma, R., \& Narayan, R. 2008, ApJ, 679, 984

\bibitem[]{}
Zhang, B., et al. 2006, ApJ, 642, 354
 
\bibitem[]{}
Zhang, B.-B., E.-W., Liang, \& Zhang, B. 2007, ApJ, 666, 1002

\bibitem[]{}
Zhang, B.-B., Zhang, B.,  E.-W., Liang, \& Wang, X.-Y. 2009, ApJ, 690, L10


\end{thebibliography}
\end{document}